\documentclass[twocolumn]{aastex61}
\usepackage{amsmath,amstext}
\usepackage[T1]{fontenc}

\newcommand{\ddv}{\rangle_{_{V}}}
\newcommand{\ddm}{\rangle_{_{M}}}
\newcommand{\dtHn}{_{\text{H}}}
\newcommand{\dtHi}{_{\text{H \tiny{II}}}}
\newcommand{\fesc}{f_{\text{esc}}}
\usepackage{float}

\shorttitle{SCORCH. II. Escape fraction}
\shortauthors{A. Doussot, H. Trac, R. Cen}

\begin{document}

\title{SCORCH. II. Radiation-Hydrodynamic simulations of reionization with varying radiation escape fractions}

\author{Aristide Doussot}
\affiliation{Ecole Normale Superieure, Departement de Physique, 24 rue Lhomond, 75005 Paris, France}
\affiliation{McWilliams Center for Cosmology, Department of Physics, Carnegie Mellon University, Pittsburgh, PA 15213}
\author{Hy Trac}
\affiliation{McWilliams Center for Cosmology, Department of Physics, Carnegie Mellon University, Pittsburgh, PA 15213}
\author{Renyue Cen}
\affiliation{Department of Astrophysical Sciences, Princeton University, Princeton, NJ 08544}

\begin{abstract}
	In the Simulations and Constructions of the Reionization of Cosmic Hydrogen (SCORCH) project, we present new radiation-hydrodynamic simulations with updated high-redshift galaxy populations and varying radiation escape fractions. The simulations are designed to have fixed Thomson optical depth $\tau \approx 0.06$, consistent with recent Planck observations, and similar midpoints of reionization $7.5 \lesssim z \lesssim 8.0$, but with different ionization histories. The galaxy luminosity functions and ionizing photon production rates in our model are in good agreement with recent HST observations. Adopting a power-law form for the radiation escape fraction $\fesc (z) = f_8[(1+z)/9]^{a_8}$, we simulate the cases for $a_8 = 0$, 1, and 2 and find $a_8 \lesssim 2$ in order to end reionization in the range $5.5 \lesssim z \lesssim 6.5$ to be consistent with Lyman alpha forest observations. At fixed $\tau$ and as the power-law slope $a_8$ increases, the reionization process starts earlier but ends later
    with a longer duration $\Delta z$ and the decreased redshift asymmetry $Az$. We find a range of durations $3.9 \lesssim \Delta z \lesssim 4.6$ that is currently in tension with the upper limit $\Delta z < 2.8$ inferred from a recent joint analysis of Planck and South Pole Telescope observations.
\end{abstract}

\keywords{cosmology: theory -  dark ages, reionization, first stars - galaxies: high-redshift  - large-scale structure of universe - methods: numerical}

\section{Introduction}

	Cosmic reionization is a major frontier topic in modern cosmology with intense ongoing theoretical and observational work. The Epoch of Reionization (EoR) starts with the formation of the first luminous sources in the first few hundred million years and ends with the reionization of hydrogen when the Universe is about a billion years old. In the process ionizing radiation from high-redshift stars, galaxies, and quasars convert the cold and neutral gas into a warm and highly ionized medium. Many fundamental questions regarding the radiation sources, reionization process, and the timing of the EoR remain. See \citet{2016ASSL..423.....M} for an excellent review.

	Recent observations from the Hubble Space Telescope (HST) and Planck have significantly improved observational constraints on the EoR. One constraint is provided 
   by the integral optical depth, measured to be $\tau = 0.058\pm 0.012$ \citep{Adam2016}. 
   Analysis suggests that, using this measurement values, the duration of the reionization is $\Delta z_{\text{CMB}}<2.8$ \citep{Adam2016}, assuming that the reionization is completed at redshift $z\approx 6$. 
   Many studies \citep[e.g.][]{Bouwens2015, Finkelstein2014, Livermore2016} show that it is likely that the main contributors to the reionization are the dwarf galaxies. 

	The project SCORCH (Simulations and Constructions of the Reionization of Cosmic Hydrogen) is motivated
    and designed to gain a deeper understanding of the EoR 
    by providing theoretical tools to improve the comparison between observations and theory. 
    To make significant progress, a systematic framework to investigate the effects of distribution and properties of radiation sources and sinks on the reionization process is needed. 
    

	Among the parameters driving the reionization, the galaxy luminosity function (GLF) is one of the less constrained at high redshift. In SCORCH I \citep{Trac2015}, we propose a way to extrapolate the known luminosity function to higher redshift and fainter magnitude. The obtained galaxy luminosity function is consistent with the measurements from \citet{Bouwens2015} and \citet{Finkelstein2014} at redshift $6\lesssim z\lesssim 10$ and also in good agreement with cosmological simulations \citep[e.g.][]{Gnedin2016,Feng2016,Liu2016,Ocvirk2016} and semi-analytical models \citep[e.g.][]{Mashian2015, Mason2015}.

	The escape fraction of ionizing photons $\fesc$, the fraction of ionizing photons which escapes from the galaxies where they have been created into the intergalactic medium,
    is a very important parameter but
    nearly impossible to observe directly at EoR.
    Overall, $\fesc$ is governed by all the internal processes of emission and absorption of ionizing photons in galaxies and relies on high resolution hydrodynamic simulations to place our understanding on a solid physical basis.
    While theoretical progress on this front is onging,
    the computed $\fesc$ can be 
    sensitively dependent on simulation methods, resolution and treatments of physical processes (such as supernova feedback), comparison among different works can be difficult currently. However, empirically, there are some constraints by assuming a given GLF with a low-luminosity limiting magnitude $M_{SF}\sim -10.0$ and a galaxy-driven reionization model, $\fesc\gtrsim 0.10-0.20$ at $z\gtrsim 6$ \citep{Bolton2007}. Curiously, at low-redshift the values obtained from observational measurements are $\fesc\lesssim 0.05-0.1$ \citep{Chen2007, Iwata2009, Smith2015} which suggests or requires that $\fesc$ varies with redshift \citep{Alvarez2012,Sun2016,Price2016}. In \citet{Price2016} both parametric and non-parametric functional forms have been tested to obtain an expression of $\fesc$ as a function of redshift that is conformal with this expected trend. It is found that $\fesc$ can be well-fitted by a simple power-law form whose exponent depends on the value of the optical depth $\tau$ that is taken as a reference.

Here, in Paper II of the SCORCH project, we produce and analyze new reionization simulations, combining previous work on the abundance of high-redshift galaxies \citep{Trac2015} and the evolution of the radiation escape fraction \citep{Price2016}. Section \ref{sec_model} describes the methods, including the radiation-hydrodynamic simulations, galaxy population models, and radiation escape fraction models. Section \ref{sec_result} presents initial results on the photoionization and photoheating. More detailed results will be presented in upcoming papers. Section \ref{sec_ccl} summarizes our work and the Appendix includes additional tests and results. We adopt the concordance cosmological parameters: $\Omega_{m}=0.30$, $\Omega_{\Lambda}=0.70$, $\Omega_{b}=0.045$, $h=0.7$, $n_{s}=0.96$, and $\sigma_{8}=0.8$

\section{Methods}\label{sec_model}

\subsection{Radiation-Hydrodynamic Simulations}

We run three new radiation-hydrodynamic simulations that are consistent with the latest observations. The simulations are designed to have fixed Thomson optical depth $\tau \approx 0.06$, consistent with recent Planck observations \citep{PlanckCollaboration2015, Adam2016, PlanckCollaboration2016}. They start with the same initial conditions, but have different reionization histories. They have the same modeled galaxy populations, but use different radiation escape fraction models. In the three simulations, $f_\text{esc}(z)$ is either constant or varies linearly or quadratically with respect to $1+z$.

	To run our simulations, we use the Radhydro code which has already been used to model both hydrogen and helium reionization \citep{Trac2008, Battaglia2013, LaPlante2016a}. In order to simultaneously solve collisional gas dynamics, collisionless dark mark dynamics, and radiative transfer of ionizing photos, the Radhydro code combines hydrodynamic and N-body algorithms \citep{Trac2003} with an adaptive ray-tracing algorithm \citep{Trac2007}. As the ray-tracing algorithm has adaptive splitting and merging, it improves the resolution and the scaling.
	
	The three Radhydro simulations, all starting with the same initial conditions at redshift $z=300$ and having $2048^3$ dark matter particles, $2048^3$ gas cells, and up to 12 billion adaptive rays. We use a fixed grid and a comoving box of side length $50\ h^{-1}$Mpc, focusing on atomic cooling halos. Consequently, we have a resolution of $24.4\ h^{-1}$kpc. For each ray we track five frequencies (15.7, 21.0, 29.6, 42.9, 74.1 eV) above the hydrogen ionizing threshold of $13.6$ eV. The two first frequencies are chosen to be below the first helium ionizing threshold, the two following frequencies are below the second helium ionizing threshold and the last frequencies is above all threshold. The nonequilibrium solvers for the ionization and energy equations use the photoionization and photoheating rates computed from the incident radiation flux. The three simulations are run down to redshift $z=$ 5.5.

	The generation of the halo and galaxy catalogs is done by a particle-particle-particle-mesh \citep[P$^3$M;][]{Trac2015} N-body simulation with $3072^3$ dark matter particles using a high-resolution version of the same initial conditions as the Radhydro simulations. Every 20 million cosmic years, a hybrid halo finder is run on the fly to locate dark matter halos and build merger trees. The particle mass resolution of $3.59 \times 10^5\ h^{-1}$M$_\odot$ allows the measurement of halo quantities such as mass and accretion rate down to the atomic cooling limit ($T \sim 10^4$ K, $M \sim 10^8\ h^{-1}$M$_\odot$). The halo mass accretion rate is calculated as
\begin{equation}
	\dot{M}=\frac{M_{2}-M_{1}}{t_{2}-t_{1}}
\label{Mass_accretion_rate_equation}
\end{equation}	
where $M_{2}$ is the mass of a given progenitor at a given time $t_{2}$ and $M_{1}$ is the mass of its descendant at a previous time $t_{1}$.
	
	The radiation sources are modeled using an updated subgrid approach allowing us to populate dark matter halos with galaxies, by matching the observed galaxy luminosity functions, while being able to 
    compute accurately the spatial distribution of ionizing sources. Starting from the halo mass accretion rate, we infer the luminosity-accretion rate relation $L_{\text{UV}}(\dot{M},z)$ from the abundance matching performed by equating the number density of galaxies to the number density of halos:
\begin{equation}
	n_\text{gal}(>L_\text{UV},z) = n_\text{halo}(>d\dot{M},z) .
\label{differential_abundance_matching}
\end{equation}
Using the halo mass accretion rate instead of the halo mass $M$ allows us to account for the scatter in mass-to-light ratio and the episodic nature of star formation. More details of the abundance matching technique can be found in SCORCH I \citep{Trac2015} and a review on reionization simulations has been done in \citet{Trac2009}.

\subsection{Galaxy Luminosity Functions}
	
	The reionization history depends strongly on the abundance of escaped ionizing photons and its evolution. The current observable galaxies with $M_{\text{UV}}\lesssim -17$ and at $z\lesssim 10$ are only a part of the ionizing sources that are responsible for the ionization history. To compute the GLF, it is therefore required to extrapolate the known luminosity function to fainter magnitude and higher redshift. We then use the fiducial model that has been created and detailed in SCORCH I \citep{Trac2015}.

\begin{figure}[t]
\centering
\includegraphics[width=0.45\textwidth]{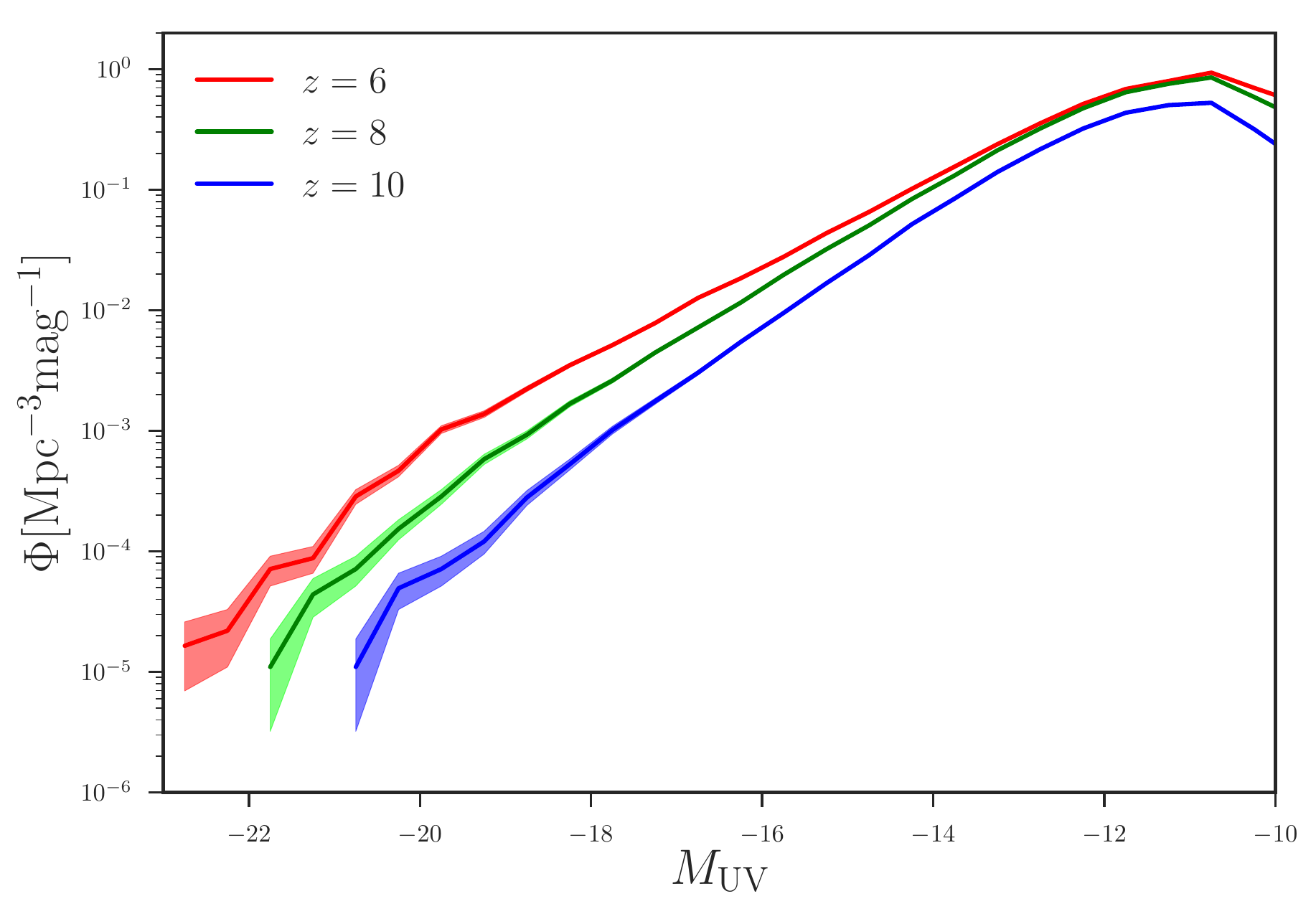}
\caption{Galaxy luminosity function (binned) as a function of the UV magnitude at $z\approx 6$ (red), $z\approx 8$ (green), and $z\approx 10$ (blue) as obtained during the SCORCH simulation.}
\label{GLF}
\end{figure}	

\begin{figure*}[!ht]
\centering
\includegraphics[width=1.0\textwidth]{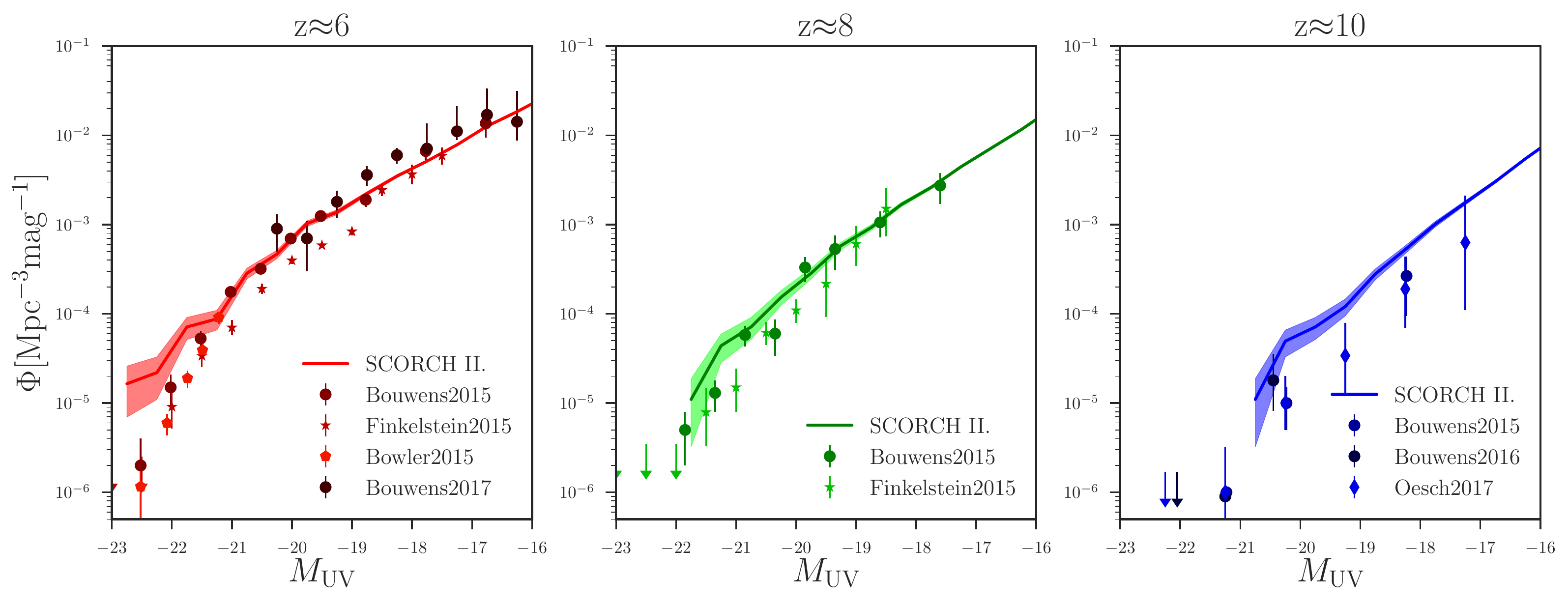}
\caption{Galaxy luminosity function (binned) as a function of the UV magnitude at $z\approx 6$ (red), $z\approx 8$ (green), and $z\approx 10$ (blue) as obtained for the SCORCH simulations. For comparison, we show the binned observational measurements at $z\approx 6$ \citep{Finkelstein2014, Bowler2015, Bouwens2015, Bouwens2016}, at $z\approx 8$ \citep{Finkelstein2014, Bouwens2015}, and  at $z\approx 10$ \citep{Bouwens2015, Bouwens2015a, Oesch2017}. For $z\approx 6$ and $z\approx 8$ our results are in agreement with the observational values and are still consistent with the uncertain observations for $z\approx 10$.}
\label{GLF_3plot}
\end{figure*}

	Figure \ref{GLF} presents the overall behaviour of our galaxy luminosity functions for the redshifts $z\approx 6, 8$, and $10$. The turn-over at fainter magnitude $M_{\text{UV}}\gtrsim -11$ has been determined for this fiducial model in \citet{Trac2015}. This limit corresponds to the minimum mass for a dark matter halo to host a galaxy assuming that galaxies are formed only in halos where the gas cools efficiently through atomic transitions.
 
	The galaxy luminosity functions for $z\approx 6,8$, and $10$ as obtained during the simulation are shown in Figure \ref{GLF_3plot}. We truncate the curves before the star formation limit and also show the observational measurements at $z\approx 6$ \citep{Finkelstein2014, Bowler2015, Bouwens2015, Bouwens2016}, at $z\approx 8$ \citep{Finkelstein2014, Bouwens2015}, and  at $z\approx 10$ \citep{Bouwens2015, Bouwens2015a, Oesch2017}. The GLF appears to be consistent with the observational constraints at $z\approx 6$ and $z\approx 8$. In the case of $z\approx 10$, our luminosity function is still consistent with the observational results but has a larger amplitude at low $M_{\text{UV}}$ than expected from the observational measurements. However the uncertainties on the measurements at $z\approx 10$ are high and our result matches the theoretical expectation from SCORCH I \citep{Trac2015}.

\subsection{Radiation Escape Fractions}

	The escape fraction of ionizing photons $\fesc$ may be computed using high resolution radiation hydrodynamic simulations \citep[e.g.,][]{2009Wise, 2014Kimm, Yajima2014, Ma2015, Kimm2016, Trebitsch2017}, although significant uncertainty remains. However, in our simulation the resolution of $24.4\ h^{-1}$kpc is insufficient to properly resolve the interstellar medium (ISM) and the circumgalactic medium (CGM) needed to self-consistently model it. We then use a parametric approach to model the escape fraction $\fesc$.
 
    In \citet{Price2016} we found that the new estimations of $\tau$ \citep{PlanckCollaboration2015, Adam2016, PlanckCollaboration2016} implies a generic redshift evolution in the radiation escape fraction $\fesc(z)$. Moreover, a simple parametric form can be used to fit that evolution. Following these results, we then chose a two-parameter single power-law
\begin{equation}
	\fesc(z)=f_{8}\left( \frac{1+z}{9}\right)^{ a_{8}}
\label{Fesc_formula}
\end{equation}
where $f_{8}$ is the value of the escape function at $z=8$. In our study, we compare the cases where $ a_{8} =0,1,$ and $2$ mainly to have a better understanding on the effect of that escape fraction on the reionization history. We chose to use Equation \ref{Fesc_formula} for all the galaxies independently of their masses. Indeed, the evolution of the escape fraction as a function of the mass of the galaxy is still uncertain, the existence of a correlation even being recently questioned \citep{Yajima2014, Ma2015}.

	Recent measurements of the Thomson optical depth yield $\tau = 0.058\pm 0.012$ \citep{Adam2016} and $\tau = 0.054\pm 0.007$ \citep{PlanckCollaboration2018}. These values imply that the universe must be half-ionized at $z\approx 8$ and by assuming that the reionization ends before $z\approx 5.5$, the exponent $ a_{8}$ in Equation \ref{Fesc_formula} is likely to be $0\lesssim a_{8}\lesssim 2$.
 
\begin{figure}[t]
\begin{center}
	\includegraphics[width=0.45\textwidth]{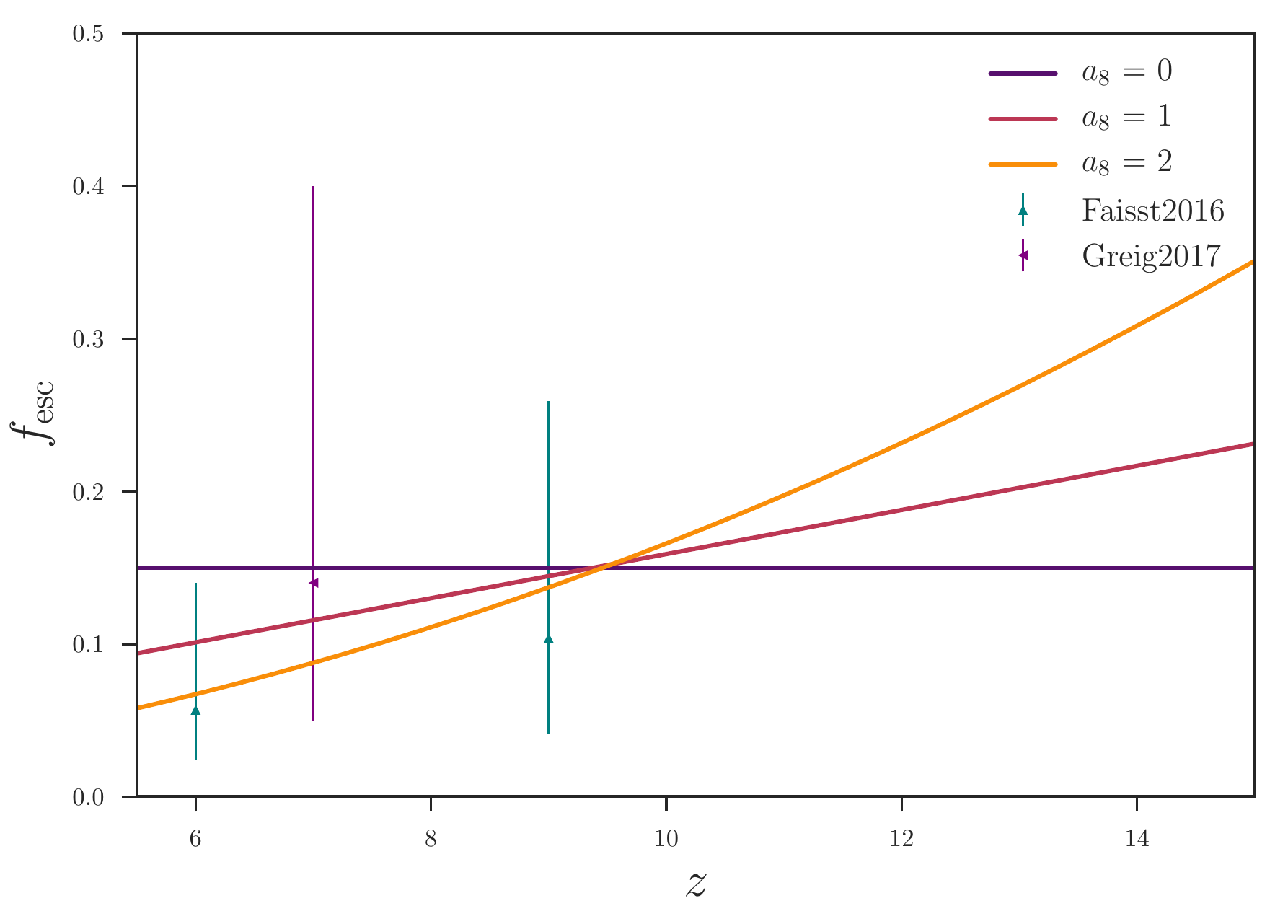}
	\caption{Escape fraction as a function of redshift following the Equation \ref{Fesc_formula} for our three models. The observationally-based predictions of \citet{Faisst2016} and \citet{Greig2017} are shown for comparison.}
	\label{Fesc}
\end{center}
\end{figure}	

 Figure \ref{Fesc} shows the behaviour of $\fesc$ as a function of the redshift for three typical values of $ a_{8}$. The parameter $f_{8}$ have been selected for the models to eventually match the value $\tau\approx 0.06$ which we choose to respect the two measurements. For $a_{8}=0,1$ and $2$, we have respectively $f_{8}=0.111, 0.130$ and $0.150$. We also compare our assumed evolutions to two recent results. The first result is from \citet{Faisst2016}, which obtain $\fesc (z=6)= 0.057^{+0.083}_{-0.033}$ and $\fesc (z=9)= 0.104^{+0.155}_{-0.063}$ for galaxies with $log(M/M_{\odot})\sim 9.0$, based on an empirical prediction of $\fesc$ made by combining the relation between $[\text{O}_{\text{ III}}/\text{O}_{\text{ II}}]$ and $\fesc$ with the redshift evolution of $[\text{O}_{\text{ III}}/\text{O}_{\text{ II}}]$ as predicted from local high-$z$ analogs. The second result is from \citet{Greig2017}, which obtain the constraint $\fesc (z\sim 7)= 0.14^{+0.26}_{-0.09}$ derived from a Bayesian framework which includes model-dependent priors from high-$z$ galaxy observations using recent observations of $z\sim$ 7 faint, lensed galaxies.

	Our three models are consistent with the latest predictions from \citet{Faisst2016} and are in good agreement with \citet{Greig2017}. Moreover, the overall normalization of $\fesc$ is degenerate with other parameters such as the overall normalization of the GLF and galaxy spectral energy distributions (SED) which may lead to some differences between different studies. Our models then broadly respect the expected escape fraction profile.

\subsection{Ionizing Photons}

	The photoionization rate $\dot{n}_{\gamma}(z)$, or the cumulative ionizing photon number density $n_{\gamma}(>z)$, is computed using the fraction of photons which have escaped from their original galaxy. The photoionization rate is then related to $\fesc$ through :
\begin{equation}
	\dot{n}_{\gamma}(z)=\fesc\times \dot{n}_{\gamma,\text{total}}(z).
\end{equation}
	For our approach we use the following formula for the production rate of ionizing photon of Population II star \citep{Trac2015}:
\begin{equation}
	\dot{N}_{\gamma}\approx 10^{46.2-0.4M_{\text{UV}}}s^{-1}\approx 10^{25.5}s^{-1}\left(\frac{L_{UV}}{\text{erg }\text{s}^{-1}\text{Hz}^{-1}} \right)
\label{photon_rate}
\end{equation}
	where the conversion between UV magnitude and luminosity has been made with the standard AB relation,
\begin{equation}
	M_{\text{UV}}=-2.5~\text{log} \left(\frac{L_{UV}}{4.345\times 10^{20} \text{ erg }\text{s}^{-1}\text{Hz}^{-1}} \right)
\label{AB_relation}
\end{equation}

	It is worth noting that Equation \ref{photon_rate} can be different from the one used in other works \citep{Trac2015} because of the normalization which is uncertain to a factor of approximatively $2$. Again, it emphasizes that we can only carefully compare our functional form of $\fesc$ with some observational constraints as these constraints are derived from a different computation of $\dot{n}_{\gamma}(z)$. However, we hereby confront our results for $\dot{n}_{\gamma}(z)$ and its cumulative $n_{\gamma}(>z)$ to another work from \citet{Bouwens2015b} as they do not depend on an arbitrary choice in their definition.
	
\begin{figure*}[!ht]
\includegraphics[width=0.5\textwidth]{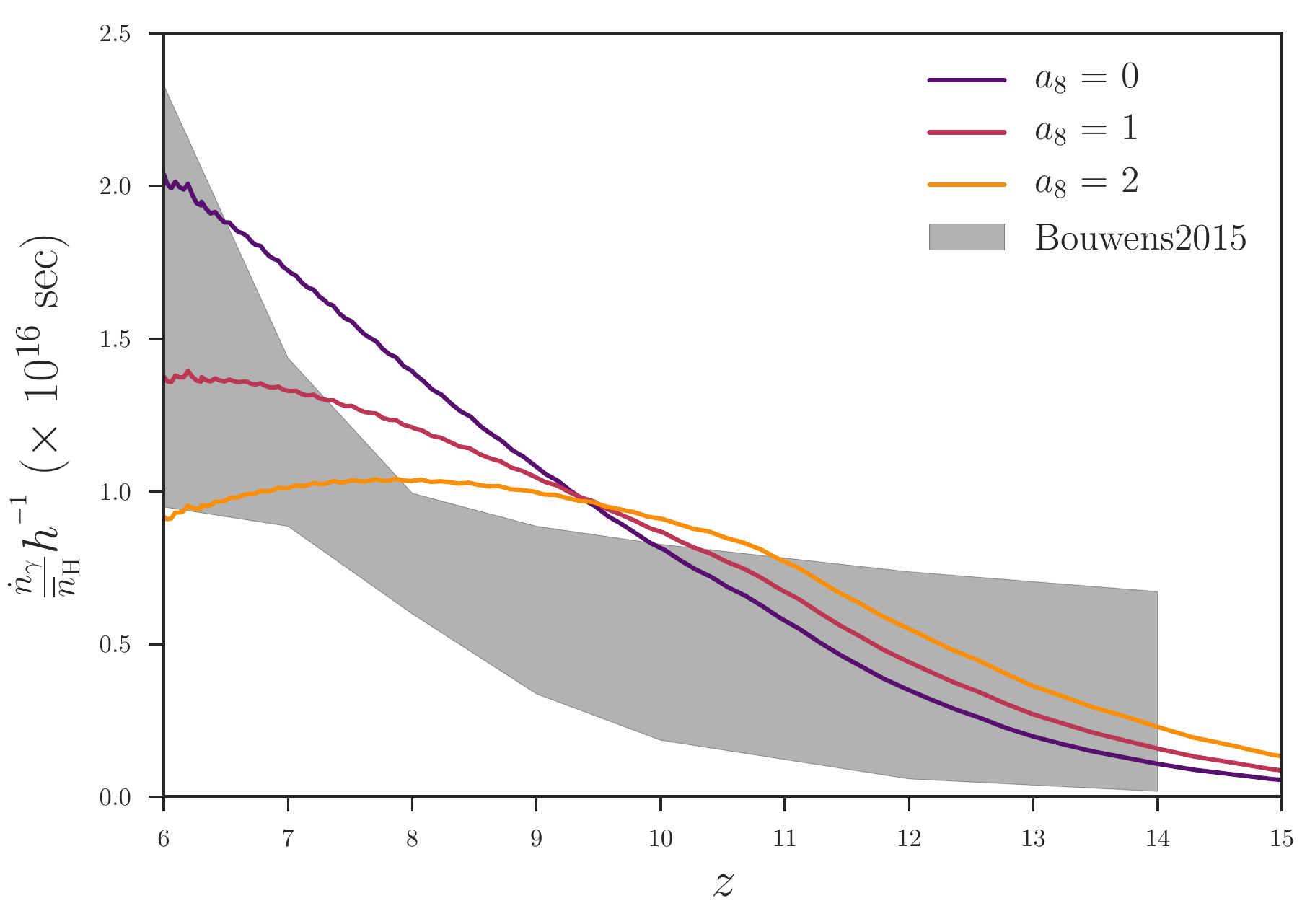}
\includegraphics[width=0.5\textwidth]{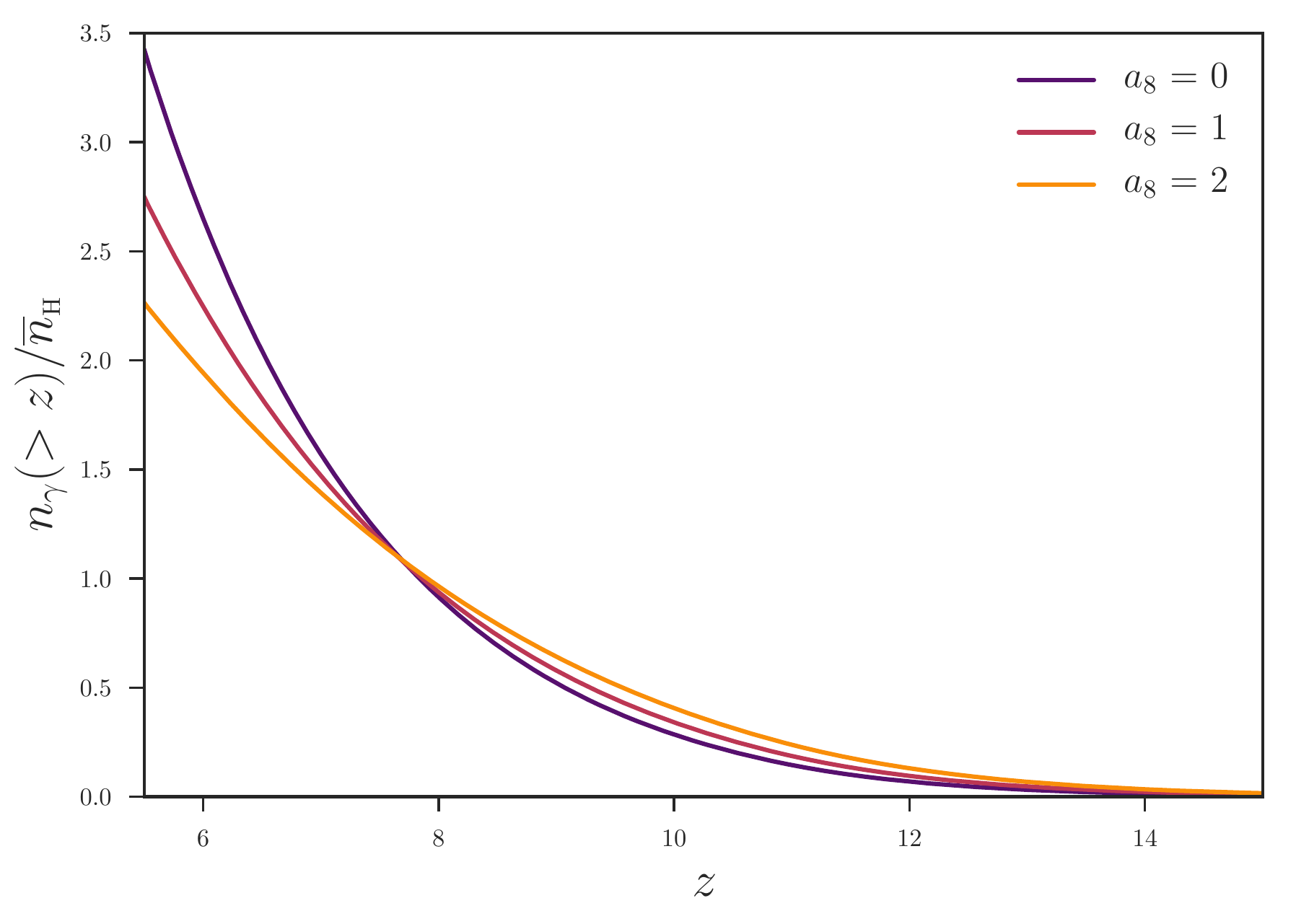}
\caption{Photoionization rate per hydrogen atoms (left) and cumulative ionizing photon number density per hydrogen atoms (right) as functions of the redshift z for our three functional forms of $\fesc$ with $ a_{8}=0$ (red), $ a_{8}=1$ (blue), and $ a_{8}=2$ (green). We show for comparison the observational results, based on HST observations, from \citet{Bouwens2015b} (shaded grey)  which considers a constant clumping factor.}
\label{N_cumul}
\end{figure*}	

	In Figure \ref{N_cumul}, we show the evolution of the photoionization rate per hydrogen atoms and of the cumulative ionizing photon number density per hydrogen atoms as functions of the redshift $z$ for our three models of escape fraction. We show that our results differ with the one from \citet{Bouwens2015b} where a constant clumping factor of $3$ is considered. It may be due to the fact that, in our study the clumping factor is not fixed and varies with the redshift resulting in a clumping factor always greater than $3$. Moreover, the photoionization rate from \citet{Bouwens2015b} does not come from a simulation but from an analytical computation, a calculation that we have also done with our varying clumping factor in Doussot et al. (in prep).

\section{Results}\label{sec_result}
\subsection{Optical Depth}

\begin{figure}[!ht]
\centering
\includegraphics[width=0.45\textwidth]{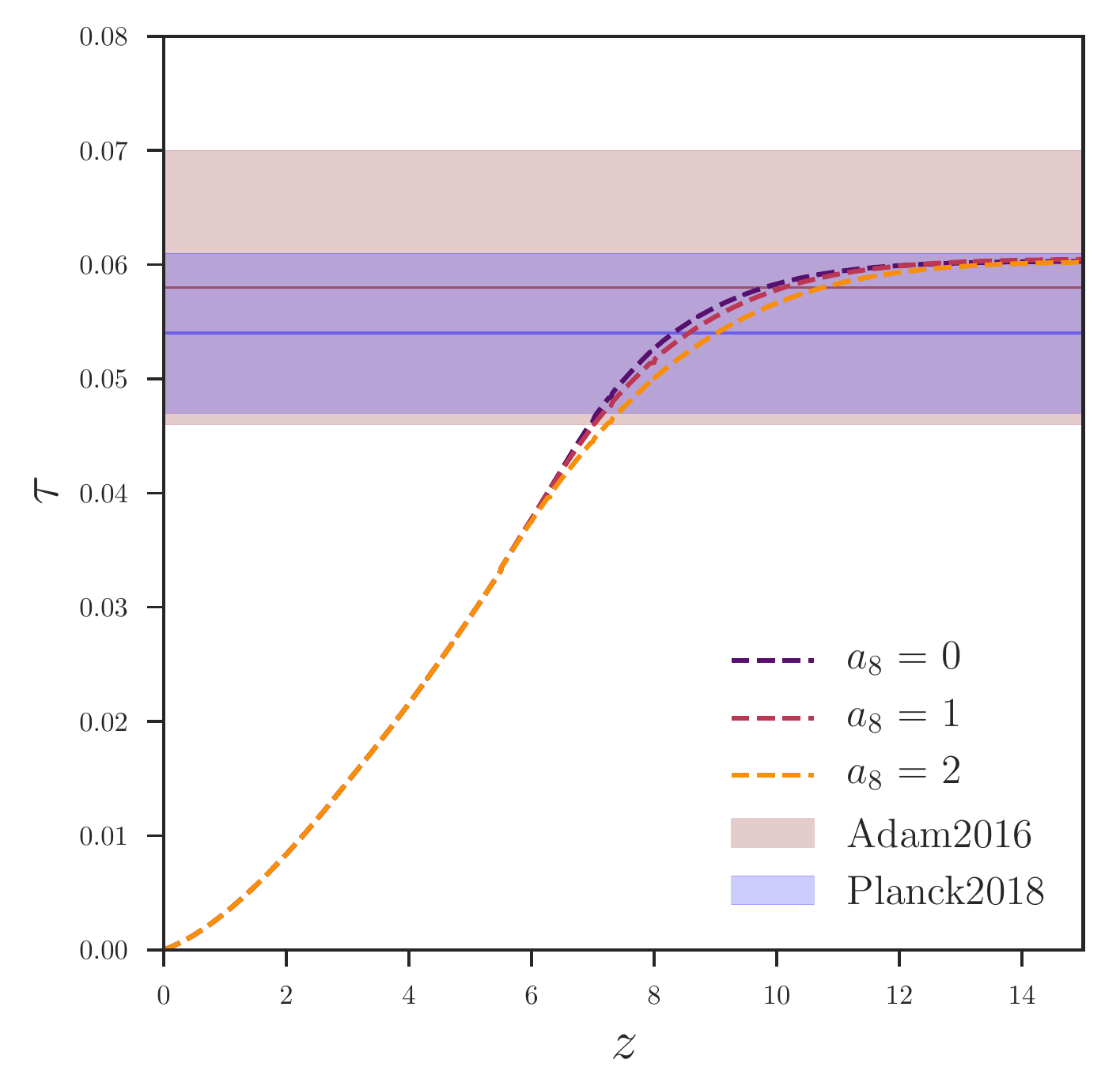}
\caption{Optical depth as a function of the redshift for our three forms of $\fesc$. The optical depth of $\tau = 0.054\pm 0.007$ from \citet{PlanckCollaboration2018} (light blue) and $\tau = 0.058\pm 0.012$ from \citet{Adam2016} (light red) are shown for comparison. The optical depths of our three models are in agreement with both measurements.}
\label{Tau}
\end{figure}	

	Figure \ref{Tau} shows the obtained optical depth $\tau$ for our three models as a function of the redshift. The observational measurements of \citet{PlanckCollaboration2018} and \citet{Adam2016}, which are respectively $\tau = 0.054\pm 0.007$ and $\tau = 0.058\pm 0.012$, are also shown for comparison. Our models were calibrated based on \citet{Adam2016} and prior to the latest results \citep{PlanckCollaboration2018}. We obtain results near $\tau\approx 0.06$, as planned by construction, and are in agreement with the measurements. However, if all of our models are consistent with the observational results, their temporal evolutions of $\tau$ are not similar which means that the reionization history is different for each one of them.

\subsection{Ionization History}

\begin{figure}[t]
\centering
\includegraphics[width=0.45\textwidth]{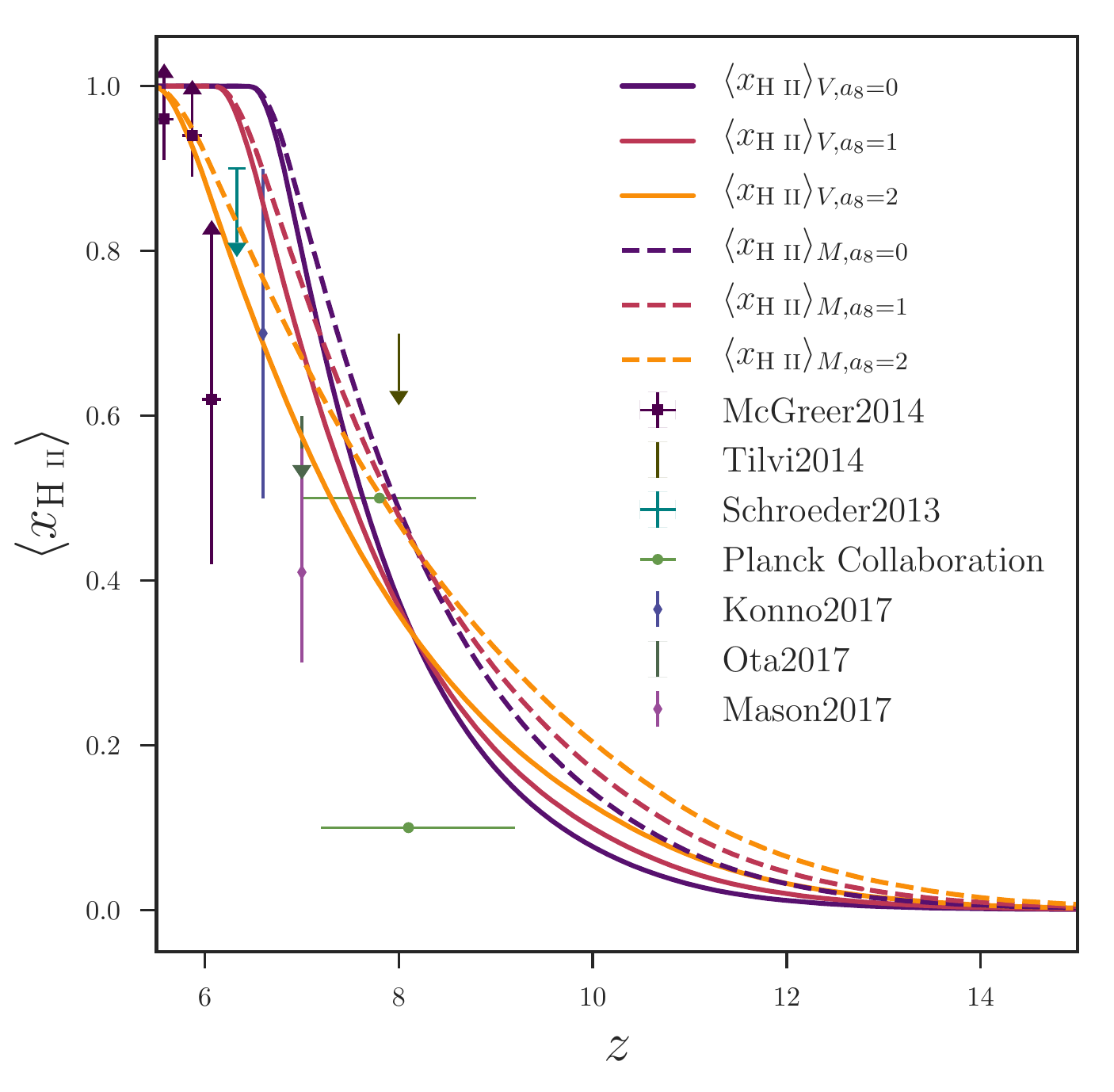}
\caption{Volume weighted (continuous) and mass weighted (dash) ionization fraction as a function of redshift for the three presented models with $z(x\dtHi=0.5)\approx 8$. We show the latest results inferred from Lyman-$\alpha$ measurements \citep{Schroeder2013, McGreer2014, Tilvi2014, Konno2017, Ota2017,Mason2017} and from Planck observations with the constraint $z(x\dtHi=0.99 )>6$ \citep{Adam2016}.}
\label{IonHist}
\end{figure}

	In Figure \ref{IonHist}, we show the volume and mass weighted ionization history, respectively $\langle x\dtHi\ddv$ and $\langle x\dtHi\ddm$, for the different studied forms of $\fesc$. We also show some of the latest results obtained from Lyman-$\alpha$ measurements \citep{Schroeder2013, McGreer2014, Tilvi2014, Konno2017, Ota2017, Mason2017} and from Planck observations with a constraint on the end of the reionization before $z\approx 6$ \citep{Adam2016}. Our results are generally in agreement with most of the experimental results cited in Figure \ref{IonHist} and no strong contradictions seem to appear.

\begin{figure*}[!ht]
\centering
\includegraphics[width=1.0\textwidth]{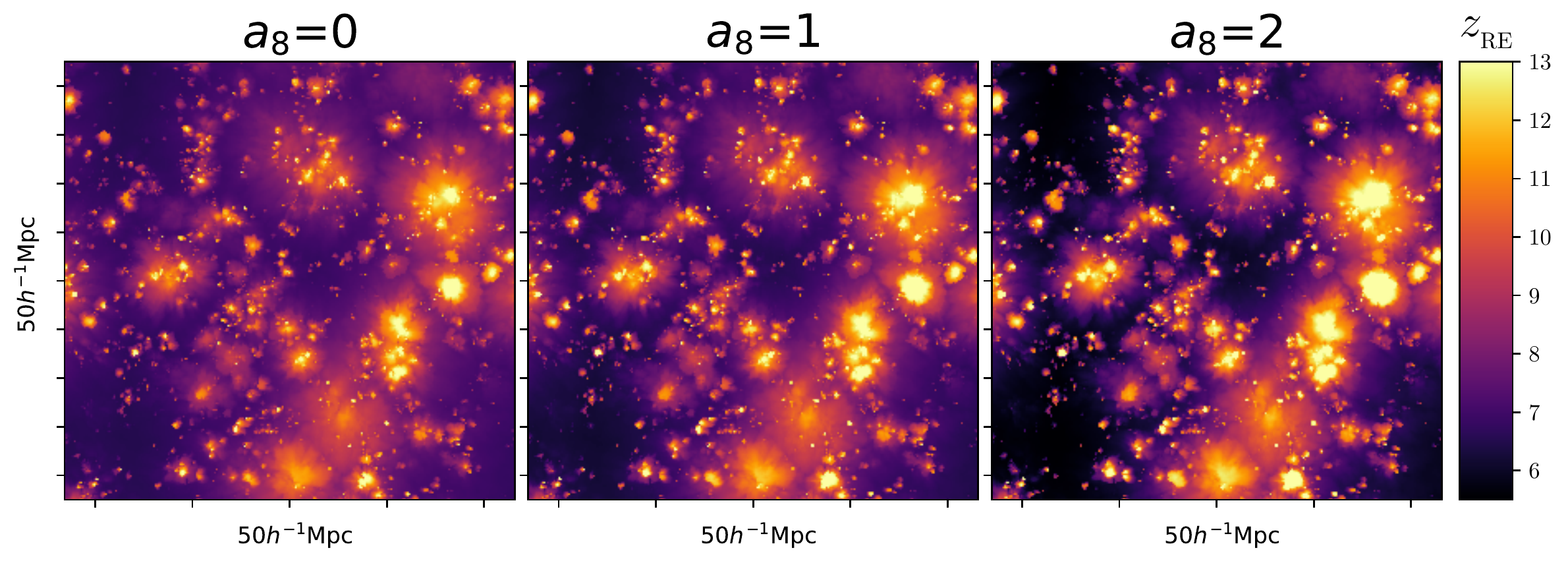}
\caption{Image of a slice of size ($50\ h^{-1}$Mpc)$^{2}$ of the reionization-redshift field (the value of one cell is the redshift when it has been ionized) for both of our three models at the end of the reionization.}
\label{Zre_2D}
\end{figure*}	

	Figure \ref{Zre_2D}, which presents slice of the reionization-redshift field for our three models, gives a first understanding of the spatial behaviour of the reionization process. We show that the reionization has the same spatial behaviour for our three models. The region around the galaxies are ionized first by the sources, the ionization grows until the whole universe is reionized and ends in the underdense cold intergalactic medium. Figure \ref{Zre_2D} also emphasizes a strong difference between the cases as the reionization in case $ a_{8} =0$ seems to have occurred in a shorter time-scale than for the two other cases whereas the reionization in the case $ a_{8} =2$ appears to extend itself in a larger time-scale.

\subsection{Duration and Asymmetry}	

	We here study how the evolution of the radiation escape fraction affects the duration of reionization. It is hard to measure from simulations the physical processes at the very beginning and the very end of the reionization. Thus, we define two durations 
\begin{equation}
	\Delta z_{50} \equiv z_{0.25}-z_{0.75}
\end{equation}
\begin{equation}
	\Delta z_{90} \equiv z_{0.05}-z_{0.95}
\end{equation}
where $z_{j}\equiv z(x\dtHi=j)$, to express the duration of the reionization like in some previous work \citep{Zahn2011, Battaglia2013}. To better characterize the reionization history, we also define the asymmetry as following:
\begin{equation}
	Az_{50}\equiv \frac{z_{0.25}-z_{0.5}}{z_{0.5}-z_{0.75}}
\end{equation}
\begin{equation}
	Az_{90}\equiv \frac{z_{0.05}-z_{0.5}}{z_{0.5}-z_{0.95}}
\end{equation}

\begin{table*}
\centering
\caption{Characterizing Values of each Model using the Mass-Weighted Ionization History $\langle x\dtHi\ddm$ and the Volume-Weighted Ionization History $\langle x\dtHi\ddv$.}
\begin{tabular}{l | c  c  c  c  c | c  c  c  c  c}
\hline \hline
 & \multicolumn{5}{c |}{$\langle x\dtHi\ddm$} & \multicolumn{5}{c}{$\langle x\dtHi\ddv$}\\
Model & $z_{0.5}$ & $\Delta z_{50}$ & $\Delta z_{90}$ & $Az_{50}$ & $Az_{90}$ & $z_{0.5}$ & $\Delta z_{50}$ & $\Delta z_{90}$ & $Az_{50}$ & $Az_{90}$\\
\hline
$ a_{8} = 0$ & 7.96 & 1.87 & 4.68 & 1.63 & 2.89 & 7.63 & 1.43 & 3.79 & 1.64 & 3.09\\
$ a_{8} = 1$ & 7.91 & 2.27 & 5.45 & 1.59 & 2.69 & 7.51 & 1.77 & 4.53 & 1.64 & 2.99\\
$ a_{8} = 2$ & 7.83 & 2.89 & 6.54 & 1.49 & 2.33 & 7.30 & 2.33 & 5.61 & 1.59 & 2.71\\
\hline
\end{tabular}
\begin{tabular}{l |  c  c  c  c  c |  c  c  c  c  c}
\hline \hline
 & \multicolumn{5}{c |}{$\langle x\dtHi\ddm$} & \multicolumn{5}{c}{$\langle x\dtHi\ddv$}\\
Model & $t_{0.5}$ & $\Delta t_{50}$ & $\Delta t_{90}$ & $At_{50}$ & $At_{90}$ & $t_{0.5}$ & $\Delta t_{50}$ & $\Delta t_{90}$ & $At_{50}$ & $At_{90}$\\
\hline
$ a_{8} = 0$ & 6.88 & 2.06 & 4.34 & 1.26 & 1.61 & 7.27 & 1.74 & 3.90 & 1.34 & 1.88\\
$ a_{8} = 1$ & 6.93 & 2.53 & 5.12 & 1.17 & 1.36 & 7.43 & 2.21 & 4.72 & 1.27 & 1.65\\
$ a_{8} = 2$ & 7.71 & 3.31 & 6.50 & 1.00 & 1.02 & 7.71 & 3.09 & 6.21 & 1.13 & 1.28\\
\hline
\end{tabular}
\tablecomments{For the second table, $t_{0.5}$, $\Delta t_{50}$, and $\Delta t_{90}$ are expressed in $10^{8}$yr.}
\label{TableHistReio}
\end{table*}

	Table \ref{TableHistReio} present, using the mass-weighted and volume-weighted ionization history, the characterizing values of each model which emphasize the differences between their reionization histories. We see that the duration of the reionization is longer when the exponent $a_{8}$ in the power-law fit of $\fesc$ is larger. Moreover, these values suggest a simple relation for $\Delta z(a_{8})$ and $Az(a_{8})$ even if other simulations with $0\lesssim a_{8}\lesssim 2$ are needed to determine it. For completeness, the trend in the evolution of the parameters is represented in Appendix \ref{AppA}. See \citet{Trac2018} for how the redshift midpoint, duration and asymmetry can be used to parametrize the reionization history.
	
	From Table \ref{TableHistReio} we also observe that, in all three cases, there is a non-negligible asymmetry on the duration of the part of the reionization before and after $x\dtHi=0.5$. The asymmetry is even larger when we consider a larger interval of the reionization history (i.e. $0.05\leq x\dtHi\leq 0.95$ instead of $0.25\leq x\dtHi\leq 0.75$). For our three simulations we then have that the first-half of the reionization ($x\dtHi<0.5$) is longer than the latter-half ($x\dtHi>0.5$). It can be explained by the fact that the photoionization rate $\dot{n}_{\gamma}$ (shown in Figure \ref{N_cumul}) gradually increases for a decreasing redshift. Hence, at the beginning of the reionization, the photoionization rate is at its lowest values explaining why the ionization process is slow.

	We also show that, while $\Delta z$ increases for an increasing $a_{8}$, the asymmetry decreases for an increasing $ a_{8}$. It is consistent with the temporal evolution of $\fesc$ and of $\dot{n}_{\gamma}$ shown in Figures \ref{Fesc} and \ref{N_cumul} because the higher the escape fraction is at the beginning of the reionization, the greater the number of ionizing photons escaping the galaxy is and so the quicker the ionization process is. Moreover the photoionization in the model $a_{8}=2$ is higher than in the two other models at high redshift and lower at low redshift. It implies an acceleration of the reionization at the beginning and a slowdown of the process at the end.
	
	The duration of the reionization is also inferred in the Planck Collaboration \citep{Adam2016} based on a joint analysis using the South Pole Telescope (SPT) measurements of the patchy kinetic Sunyaev-Zel'dovich \citep[KSZ;][]{Sunyaev1970, Ostriker1986} effect angular power spectrum at $l$=$3000$ \citep{George2015} and only our theoretical models from \citet{Battaglia2013a}. Assuming that $z_{0.99}>6$,
\begin{equation}	
	\Delta z_{\text{CMB}}\equiv z_{0.1}-z_{0.99} < 2.8 \text{ (95$\%$ confidence)}
	\label{duration_Planck}
\end{equation}

	However this result depends on the assumptions made in the analysis and modelling of the patchy KSZ angular power spectrum. Using the definition \ref{duration_Planck} of the duration, we obtain the results presented in Table \ref{Value_duration_Planck}.
	
\begin{table}
\centering
\caption{Duration of the Reionization with the Definition \ref{duration_Planck} using Respectively the Mass-Weighted Ionization History $\langle x\dtHi\ddm$ the Volume-Weighted Ionization History $\langle x\dtHi\ddv$.}
\begin{tabular}{c  c  c  c}
\hline \hline
$a_{8}$ & $0$ & $1$ & $2$\\
\hline
$\Delta z_{\text{CMB},\langle x\dtHi\ddm}$ & 3.9 & 4.6 & 5.7\\
$\Delta z_{\text{CMB},\langle x\dtHi\ddv}$ & 3.1 & 3.8 & 4.8\\
\hline
\end{tabular}
\label{Value_duration_Planck}
\end{table}
	
	Our model $a_{8}=2$ does not match the constraint $z_{0.99}>6$ as it can be seen in Table \ref{TableSimRes} but the two other models do respect it. Accordingly, our theoretical predictions and the observational constraints from the Planck Collaboration are currently in tension. However, there are some limits to emphasize about both works. From an observational point of view, there are still large uncertainties in isolating the KSZ effect and the power spectrum from the other components like the cosmological microwave background (CMB), the thermal Sunyaev-Zel'dovich (TSZ) effect and the cosmic infrared background (CIB) using only 3 frequency bands by the South Pole Telescope. From a theoretical point of view, the patchy and homogeneous KSZ components are still imperfectly modeled. There is also discrepancies in predictions for the homogeneous KSZ effect and power spectrum \citep[e.g.][]{Trac2011, Shaw2012} implying that more work is needed to improve the simulations. Furthermore, in \citet{Adam2016}, our theoretical models from \citet{Battaglia2013a} are used to consider the homogeneous and patchy KSZ contibutions. It is important to note that the patchy KSZ effect from \citet{Battaglia2013a} are based on semi-numerical models that have only minor asymmetry compared to our current RadHydro simulations. Their patchy KSZ power spectrum are fitted based only on the value of $z_{0.5}$ and the duration of the reionization, but the asymmetry parameter is required for more precise constraints. Finally, there are some recent works that tend to show that quasars contribution to the reionization on large scale can accelerate the end of the process \citep{Madau2015, Chardin2015, DAloisio2016}. In this case the duration of the reionization is shortened while the escape fraction $\fesc$ for the galaxy population stays the same. Consequently, the two results can probably be reconciled by removing some of these previous limits.

\subsection{Temperature}

\begin{figure}[t]
\centering
\includegraphics[width=0.45\textwidth]{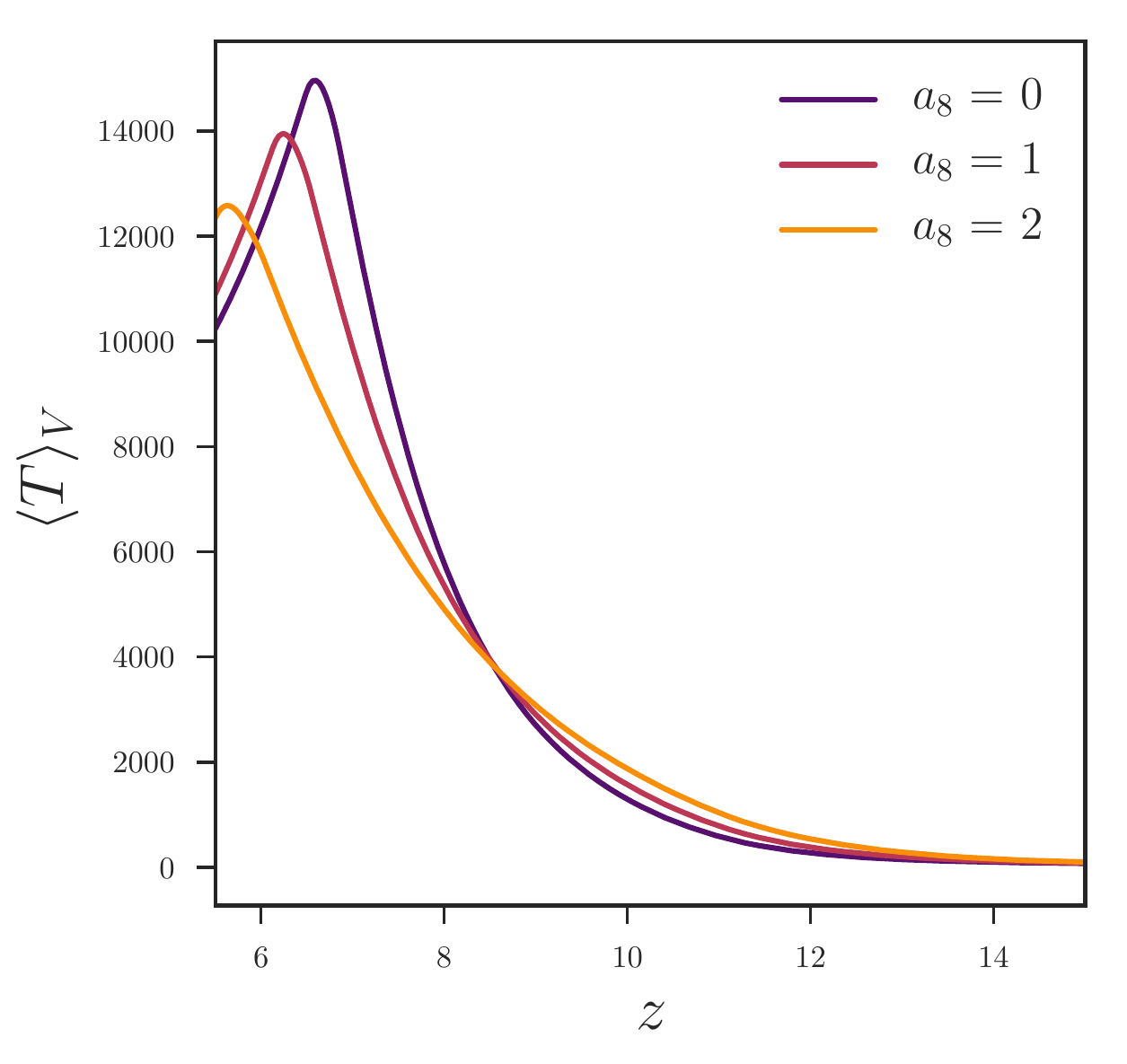}
\caption{Volume weighted average temperature $\langle T\ddv$ as a function of redshift for the three models.}
\label{Tofz}
\end{figure}	

	Like the photoionization, the photoheating is also impacted by the evolution of the radiation and its behaviour in our simulations needs to be shown for completeness. Figure \ref{Tofz} presents the volume weighted average temperature $\langle T\ddv$ as a function of redshift for the three models. At $z\gtrsim 8.5$ we can see that the photoheating process is more advanced in the model $a_{8} =2$ following the fact that the ionization fraction is higher in this case at this redshift. On the contrary, at $z\lesssim 7.5$, the temperature in the model $ a_{8} =0$ is higher than in the other models because of the fact that the ionization is higher for this model at this redshift. We also show that the maximal temperature is higher in the model $ a_{8} =0$. It is due to the shorter duration of the reionization $\Delta z$ which implies that the effect of the adiabatic cooling process during the reionization, due to the universe expansion, is smaller that it could have been on a longer period. The photo-heating history of the simulations are in agreement with their photo-ionization history highlighting the self-consistency of our results.
	
	After the end of the ionization process, in all cases, the photoheating process cannot stabilize the temperature at its higher values leading to a cooling of the gas. That cooling is in a more advanced state in the model with $a_{8} =0$ than in the other because the reionization ended up early giving to the gas more time to cool.

\begin{figure}[t]
\centering
\includegraphics[width=0.45\textwidth]{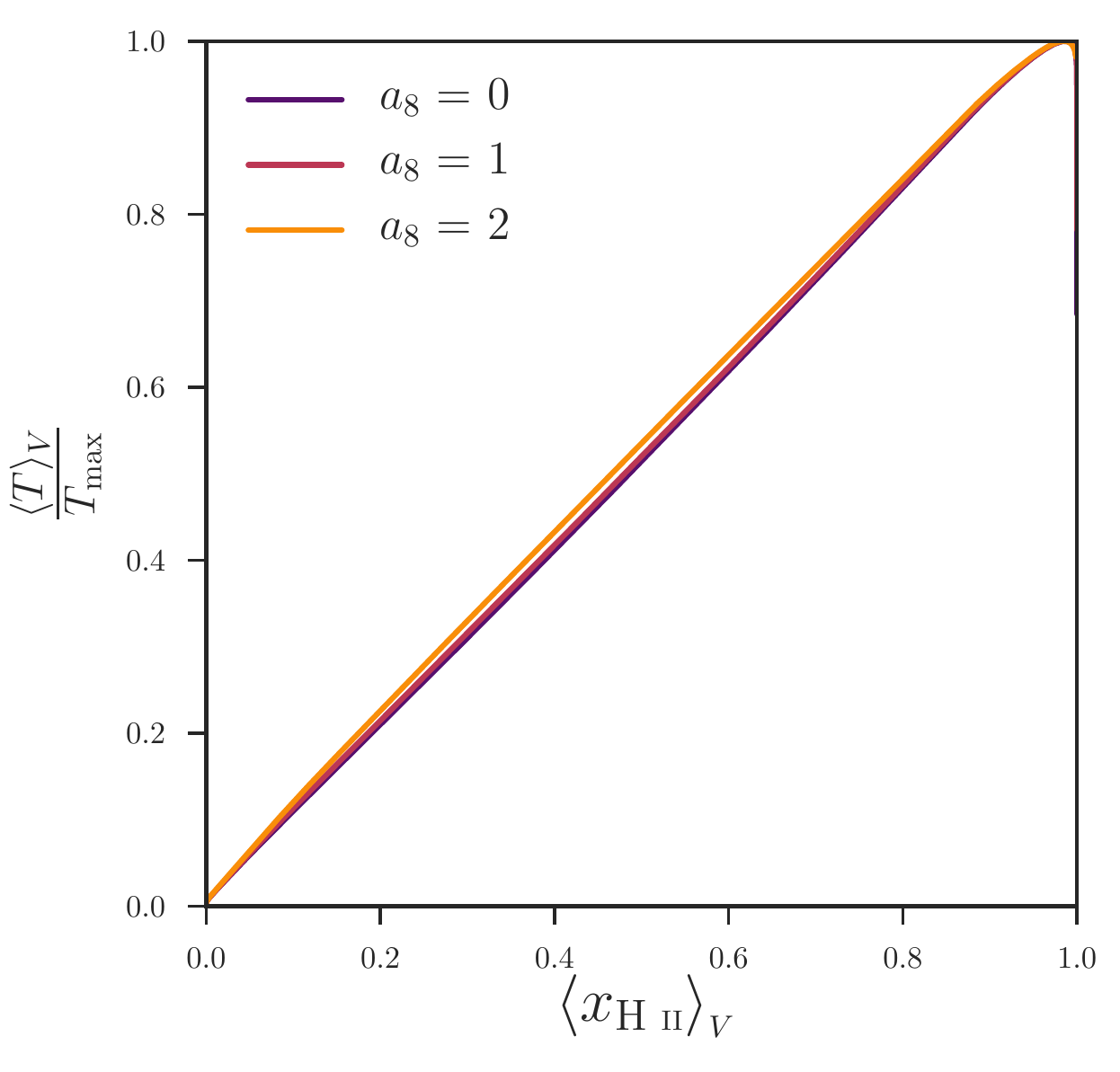}
\caption{Normalized volume weighted average temperature of the sampling box as a function of the volume weighted ionization fraction of hydrogen $x\dtHi$ for the three models.}
\label{Tofx}
\end{figure}	

	Despite their differences, the underlying physical process in all models should be the same as expected from a theoretical point of view. In Figure \ref{Tofx}, we show the normalized volume weighted average temperature of the sampling box as a function of the volume weighted ionization fraction of hydrogen $\langle x\dtHi\ddv$. Analyzing $\langle T\ddv$ as a function of $\langle x\dtHi\ddv$ and normalizing it by its maximal value allow us to cancel the influence of the different durations of the reionization $\Delta z$ of the models. As the behaviour is strictly the same for all models, we have another endorsement of the photo-heating evolution of the simulations matching the expectations. We also show a nearly linear dependency between $\langle T\ddv$ and $\langle x\dtHi\ddv$ with a coefficient of proportionality of the order of $1$.

\begin{figure*}[!ht]
\centering
\includegraphics[width=1.0\textwidth]{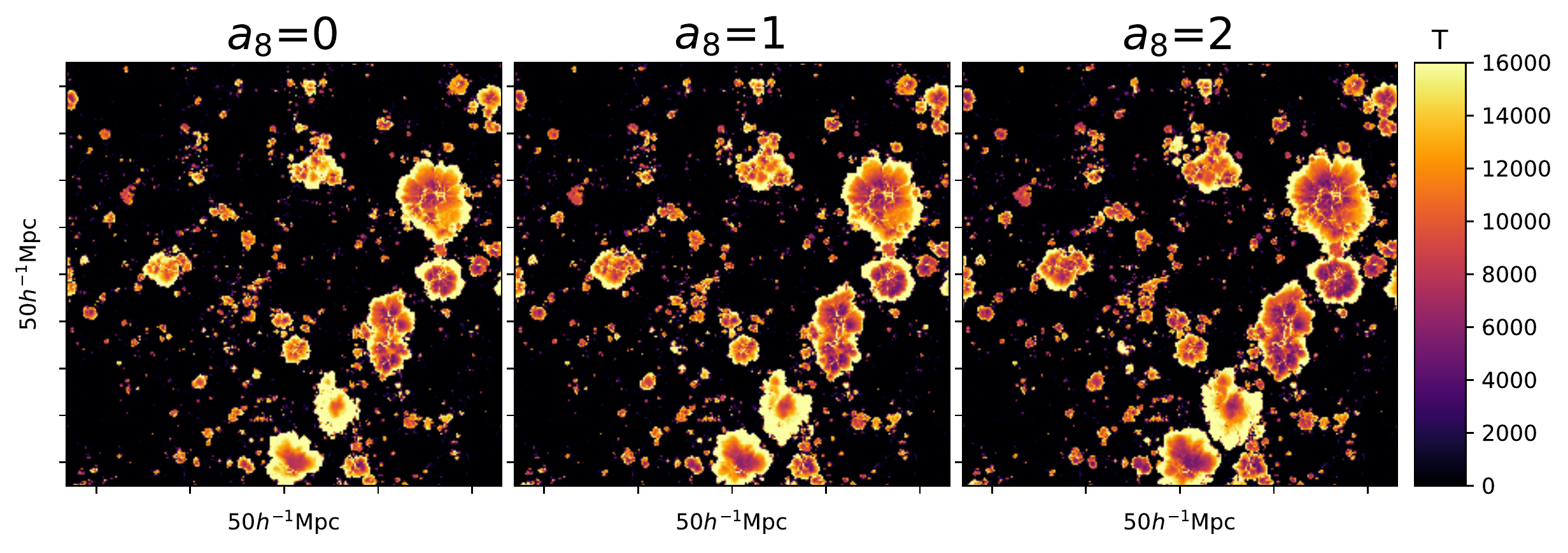}
\includegraphics[width=1.0\textwidth]{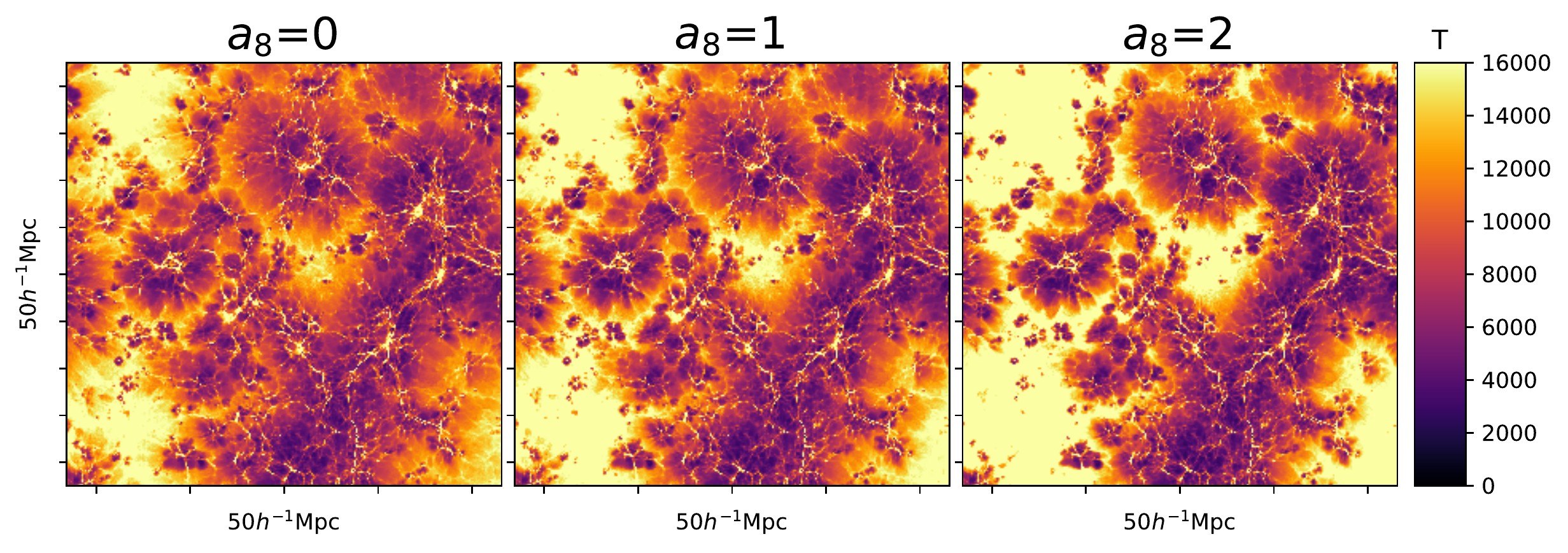}
\caption{Image of a slice of size ($50\ h^{-1}$Mpc)$^{2}$ of the temperature field for our three models at $z\approx 9.5$ (top) and $z=5.5$ (bottom).}
\label{T_2D}
\end{figure*}	

	We show in Figure \ref{T_2D} the temperature distribution of the same slice of our sampling box at the same redshift for our three different models to give a first insight of the spatial distribution of the temperature. By referring to these temperature fields, we confirm that, as expected, the temperature evolution in our simulations is strongly linked to the reionization history and so changes with the assumed functional form of the escape fraction. At $z\approx 9.5$ the increase of the temperature, revealing the position of the ionization front, is located around the sources where the reionization has started. At $z=5.5$ it is worth noting that, in all models, the gas in the vast underdense intergalactic medium is hotter than the high density gas closer to the sources. As previous work has emphasized it \citep[e.g.][]{Trac2008}, it is because the gas in the IGM has been ionized later than the gas around the sources that the former has less time to cool than the latter.
    
    A broader study of the spatial distribution of the temperature and its heterogeneity depending on the density of the gas will be done in future work. For example, see \citet{2018arXiv180709282D} for a more in depth analysis of the heating of the IGM by hydrogen reionization.

\section{Conclusion}\label{sec_ccl}

	The new RadHydro simulations based on the works of \citet{Trac2015} and \citet{Price2016} allow us to have a better understanding of the Epoch of Reionization and of the global behaviour of the parameters that constraint that epoch. In this study, we have presented the first main results of that simulation for three different cases. These cases come from the form of the escape fraction $\fesc$ which has been fitted as a simple power-law form and which has been shaped in three different ways: to be constant ($a_{8}=0$) as assumed in most previous works, to vary linearly ($a_{8}=1$), and to vary quadratically ($a_{8}=2$).
	
	Each of these models matches the observational values of the optical depth $\tau$ from \citet{Adam2016} and \citet{PlanckCollaboration2018} and are then consistent with the observations. Based on these cases, we can isolate the $\fesc$ dependency of the reionization history. We concluded that the duration of the reionization $\Delta z$ increases with the increase of the exponent of the escape fraction's power-law. On the contrary the asymmetry $Az$ between the beginning and the end of the reionization decreases with the increase of that exponent. However, our duration of the reionization conflicts with the result from \citet{Adam2016} which highlights that more studies as well as a better observation and modelization of the KSZ effect are needed.
	
	In term of the photoheating, we pointed out that the increase in temperature happens during the ionization process, and then that there is a correlation between the temperature and the reionization history. We have also shown that the maximum value of the temperature is related to the duration of the reionization $\Delta z$ a shorter duration leading to lesser time for the adiabatic cooling process to act and consequently to a harsher heating. However, by normalizing the temperatures by their maximum values and showing them as a function of the ionization fraction $x\dtHi$, it is relevant to think that the underlying photoheating process of the reionization stays the same whatever the model of escape fraction. Finally, after the ionization process, the gas starts to cool which spatially results in the fact that, at the end of the reionization, the underdense gas regions are hotter than the overdense gas regions which were ionized earlier and had more time to cool. 
\\

	We thank Anson D'Aloisio, Francois Lanusse, and Michelle Ntampaka for helpful discussions. AD acknowledges the McWilliams Center for Cosmology for hosting his internship. HT acknowledges support from NASA grant ATP-NNX14AB57G and STSCI grant HST-AR-15013.002-A. RC acknowledges support from NASA grant 80NSSC18K1101. Simulations were run at the NASA Advanced Supercomputing (NAS) Center.
	
\clearpage

\appendix
\section{Convergence test}\label{AppA}

	To study the $\fesc$ dependency of the duration of the reionization $\Delta z$ and its asymmetry $Az$, we hereby present our analysis of the convergence of these values by increasing the resolution.

\begin{figure}[h!]
\centering
\includegraphics[width=0.85\textwidth]{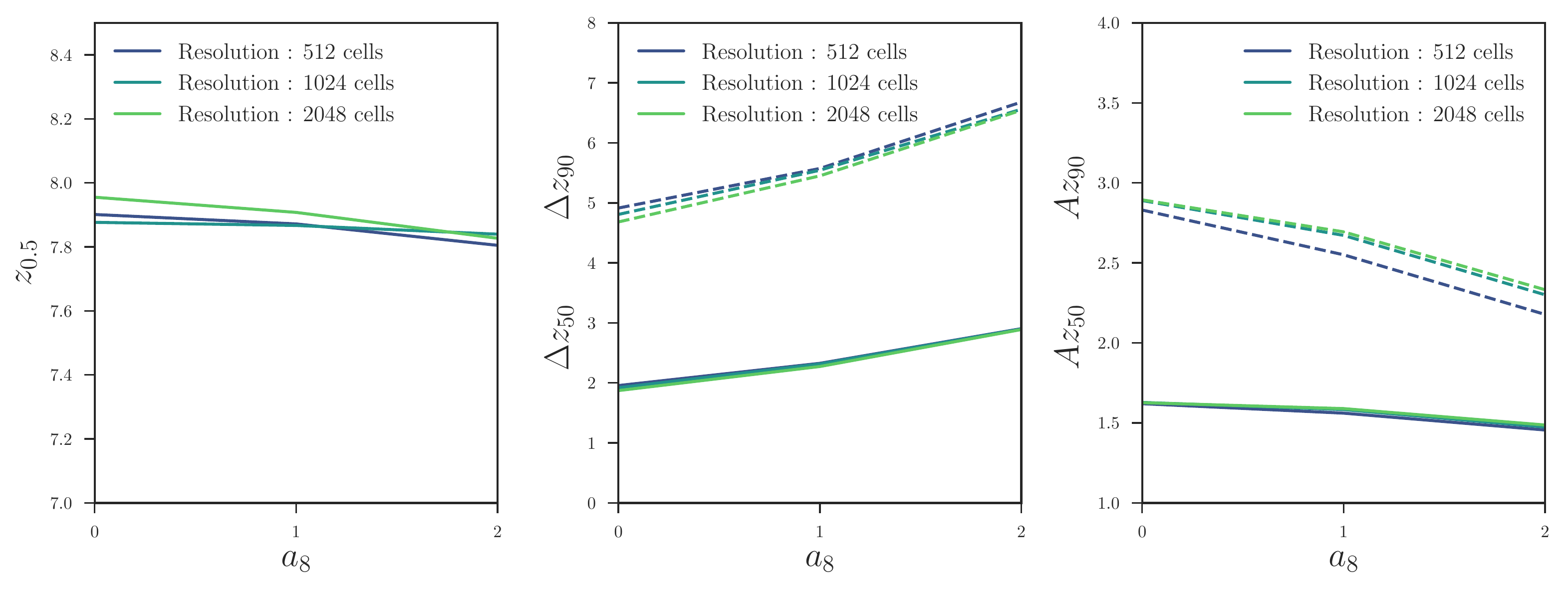}
\caption{Characterizing values of the reionization as a function of $a_{8}$ and using the mass-weighted ionization history $\langle x\dtHi\ddm$  while varying the resolution of the simulation with: at left $z_{0.5}$ the redshift at which $\langle x\dtHi\ddm =0.5$, in the middle the duration of the reionization $\Delta z_{50}$ (continuous) and $\Delta z_{90}$ (dash) and at right the asymmetry $Az_{50}$ (continuous) and $Az_{90}$ (dash).}
\label{ReioDurM}
\end{figure}	

\begin{figure}[h!]
\centering
\includegraphics[width=0.85\textwidth]{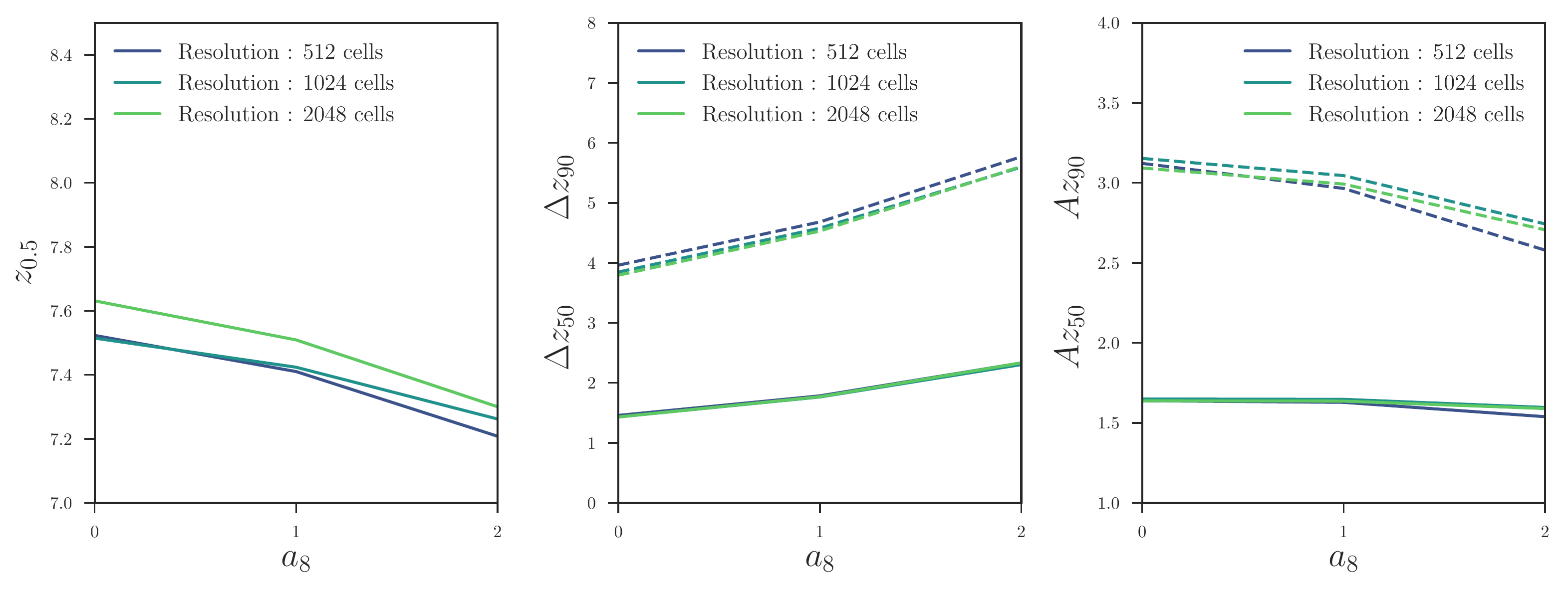}
\caption{Characterizing values of the reionization as a function of $a_{8}$ and using the volume-weighted ionization history $\langle x\dtHi\ddv$  while varying the resolution of the simulation with: at left $z_{0.5}$ the redshift at which $\langle x\dtHi\ddv =0.5$, in the middle the duration of the reionization $\Delta z_{50}$ (continuous) and $\Delta z_{90}$ (dash) and at right the asymmetry $Az_{50}$ (continuous) and $Az_{90}$ (dash).}
\label{ReioDur}
\end{figure}	

	Figures \ref{ReioDurM} and \ref{ReioDur} shows the duration of the reionization $\Delta z$ and the asymmetry of the process $Az$ depending on the resolution using respectively the mass-weighted and the volume-weighted ionization history. As we presented it above, $\Delta z$ seems to increase and $Az$ to decrease when $a_{8}$ increases. Here, we show that this conclusion does not depend on the resolution of the simulation. Moreover, especially in the case of the duration $\Delta z$, the values appear to converge to a certain limit value when the resolution increases. That fact allow us to conclude that the previously deduced dependencies are likely true and are not just a result of our limited resolution. However more studies for other values of $z_{0.5}$ and then $a_{8}$ are required to really endorse that conclusion.

\clearpage

\section{Simulation Results}

	We summarize in Table \ref{TableSimRes} the values of the main quantities obtained from SCORCH RadHydro simulation for a representative redshift set.

\begin{table}[h!]
\centering
\caption{Characteristic Quantities from SCORCH for $ a_{8} =0$, $1$, and $2$.}
\begin{tabular}{l | c  c  c | c  c  c | c  c  c}
\hline \hline
 & \multicolumn{3}{c |}{$a_{8}=0$} & \multicolumn{3}{c |}{$a_{8}=1$} & \multicolumn{3}{c}{$a_{8}=2$}\\
$z$ & $\langle x\dtHi\ddm$ & $\langle x\dtHi\ddv$ & $\frac{n_{\gamma}(>z)}{n\dtHn}$ & $\langle x\dtHi\ddm$ & $\langle x\dtHi\ddv$ & $\frac{n_{\gamma}(>z)}{n\dtHn}$ & $\langle x\dtHi\ddm$ & $\langle x\dtHi\ddv$ & $\frac{n_{\gamma}(>z)}{n\dtHn}$\\
\hline
13.5 & 0.009 & 0.002 & 0.021 & 0.014 & 0.005 & 0.032 & 0.023 & 0.01 & 0.049\\
13.25 & 0.011 & 0.003 & 0.026 & 0.018 & 0.006 & 0.039 & 0.028 & 0.012 & 0.058\\
13.0 & 0.014 & 0.004 & 0.032 & 0.022 & 0.008 & 0.047 & 0.033 & 0.015 & 0.068\\
12.75 & 0.017 & 0.005 & 0.039 & 0.026 & 0.01 & 0.056 & 0.04 & 0.018 & 0.081\\
12.5 & 0.021 & 0.007 & 0.047 & 0.032 & 0.013 & 0.067 & 0.047 & 0.022 & 0.095\\
12.25 & 0.026 & 0.009 & 0.057 & 0.038 & 0.016 & 0.08 & 0.055 & 0.027 & 0.112\\
12.0 & 0.032 & 0.012 & 0.069 & 0.046 & 0.02 & 0.096 & 0.065 & 0.032 & 0.131\\
11.75 & 0.039 & 0.015 & 0.084 & 0.055 & 0.024 & 0.113 & 0.076 & 0.039 & 0.152\\
11.5 & 0.048 & 0.019 & 0.101 & 0.065 & 0.03 & 0.134 & 0.089 & 0.046 & 0.177\\
11.25 & 0.058 & 0.024 & 0.121 & 0.078 & 0.037 & 0.158 & 0.103 & 0.055 & 0.206\\
11.0 & 0.07 & 0.031 & 0.145 & 0.092 & 0.046 & 0.187 & 0.12 & 0.067 & 0.238\\
10.75 & 0.085 & 0.04 & 0.174 & 0.109 & 0.057 & 0.219 & 0.138 & 0.079 & 0.275\\
10.5 & 0.102 & 0.05 & 0.206 & 0.128 & 0.069 & 0.255 & 0.159 & 0.093 & 0.315\\
10.25 & 0.121 & 0.062 & 0.243 & 0.148 & 0.083 & 0.296 & 0.18 & 0.109 & 0.358\\
10.0 & 0.143 & 0.077 & 0.285 & 0.171 & 0.1 & 0.341 & 0.204 & 0.127 & 0.406\\
9.75 & 0.168 & 0.095 & 0.333 & 0.197 & 0.118 & 0.391 & 0.229 & 0.146 & 0.458\\
9.5 & 0.197 & 0.116 & 0.388 & 0.225 & 0.14 & 0.447 & 0.256 & 0.168 & 0.515\\
9.25 & 0.23 & 0.141 & 0.45 & 0.257 & 0.165 & 0.51 & 0.285 & 0.191 & 0.576\\
9.0 & 0.268 & 0.171 & 0.521 & 0.292 & 0.194 & 0.579 & 0.317 & 0.218 & 0.642\\
8.75 & 0.312 & 0.209 & 0.602 & 0.331 & 0.228 & 0.655 & 0.351 & 0.248 & 0.714\\
8.5 & 0.362 & 0.254 & 0.693 & 0.375 & 0.268 & 0.74 & 0.387 & 0.281 & 0.791\\
8.25 & 0.42 & 0.309 & 0.797 & 0.424 & 0.314 & 0.834 & 0.426 & 0.318 & 0.875\\
8.0 & 0.487 & 0.375 & 0.915 & 0.478 & 0.367 & 0.938 & 0.469 & 0.359 & 0.964\\
7.75 & 0.563 & 0.456 & 1.049 & 0.539 & 0.43 & 1.052 & 0.514 & 0.404 & 1.061\\
7.5 & 0.65 & 0.553 & 1.202 & 0.606 & 0.503 & 1.179 & 0.563 & 0.456 & 1.164\\
7.25 & 0.747 & 0.667 & 1.374 & 0.679 & 0.586 & 1.318 & 0.615 & 0.512 & 1.275\\
7.0 & 0.851 & 0.798 & 1.57 & 0.76 & 0.682 & 1.471 & 0.67 & 0.574 & 1.393\\
6.75 & 0.951 & 0.931 & 1.792 & 0.845 & 0.789 & 1.64 & 0.729 & 0.643 & 1.519\\
6.5 & 0.999 & 0.998 & 2.044 & 0.93 & 0.903 & 1.824 & 0.791 & 0.718 & 1.652\\
6.25 & 1.0 & 1.0 & 2.328 & 0.989 & 0.985 & 2.026 & 0.854 & 0.799 & 1.793\\
6.0 & 1.0 & 1.0 & 2.65 & 1.0 & 1.0 & 2.247 & 0.919 & 0.886 & 1.943\\
5.75 & 1.0 & 1.0 & 3.011 & 1.0 & 1.0 & 2.485 & 0.972 & 0.96 & 2.098\\
5.5 & 1.0 & 1.0 & 3.425 & 1.0 & 1.0 & 2.75 & 0.999 & 0.999 & 2.264\\
\hline
\end{tabular}
\label{TableSimRes}
\end{table}

\clearpage

\bibliographystyle{apj}
\bibliography{CMU-Scorch_2}

\end{document}